\documentclass[12pt,preprint]{aastex}

\def\alwaysmath#1{\ifmmode{#1}\else{$#1$}\fi}

\def\ltsima{$\; \buildrel < \over \sim \;$}
\def\gtsima{$\; \buildrel > \over \sim \;$}
\def\lsim{\lower.5ex\hbox{\ltsima}}
\def\gsim{\lower.5ex\hbox{\gtsima}}
\def\lapp{\ifmmode\stackrel{<}{_{\sim}}\else$\stackrel{<}{_{\sim}}$\fi}
\def\gapp{\ifmmode\stackrel{>}{_{\sim}}\else$\stackrel{<}{_{\sim}}$\fi}
\def\rhalf{\alwaysmath{r_{1/2}^{\scriptscriptstyle\rm BSS}}} 
\def\rcore{\alwaysmath{r_{c}}}

\def\bbsshb{\alwaysmath{F_{\scriptscriptstyle\rm HB}^{\scriptscriptstyle 
   \rm bBSS}}} 
\newdimen\minuswidth    
\setbox0=\hbox{$-$}
\minuswidth=\wd0
\newdimen\digitwidth    
\setbox0=\hbox{\rm0}
\digitwidth=\wd0
\catcode`!=\active
\def!{\kern\digitwidth}
 
\shorttitle{BSS in NGC~6752} 
\shortauthors{Sabbi et al.} 
 
\begin{document} 
 
\title{The small Blue Straggler star population in the dense
Galactic Globular Cluster NGC~6752
\footnote{Based on observations with the NASA/ESA HST,
obtained at the Space Telescope Science Institute, which is operated
by AURA, Inc., under NASA contract NAS5-26555. Also based on WFI
observations collected at the European Southern Observatory,
La Silla, Chile, within the observing
programme 62.L-0354 and 64.L-0439.}
}
 
\author{E.~Sabbi\altaffilmark{1}}
\affil{\altaffilmark{1}Dipartimento di Astronomia, Universit\`a di
Bologna, via Ranzani 1,I--40126 Bologna, Italy}
\email{elena.sabbi@unibo.it}

\author{F.~R.~Ferraro\altaffilmark{1}}
\email{francesco.ferraro3@unibo.it}

\author{A.~Sills\altaffilmark{2}}
\affil{\altaffilmark{2}Department of Physics and Astronomy, McMaster
University, 1280 Main Street West, Hamilton, ON~L8S~4M1, Canada}
\email{asills@mcmaster.ca}

\and

\author{R.~T.~Rood\altaffilmark{3}} 
\affil{\altaffilmark{3}Astronomy Dept., University of Virginia
Charlottesville VA~22903--0818, USA} 
\email{rtr@virginia.edu}
 
\begin{abstract}

We have used high resolution {\it WFPC2-HST} and wide field
ground-based observations to construct a catalog of blue straggler
stars (BSS) which spans the entire radial extent of the globular
cluster NGC~6752. The BSS sample is the most extensive ever obtained
for this cluster. Though NGC~6752 is a high density cluster with a
large binary population, we found that its BSS content is surprisingly
low: the specific number of BSS is among the lowest ever measured in a
cluster. The BSS distribution is highly peaked in the cluster center,
shows a rapid decrease at intermediate radii, and finally rises again
at larger distances. This distribution closely resembles those
observed in M3 and 47Tuc by \citet{ferraro93, ferraro03c}. To date,
BSS surveys covering the central regions with HST and the outer
regions with wide field CCD ground-based observations have been
performed for only these three clusters. Despite the different
dynamical properties, a bimodal radial distribution has been found in
each. A detailed comparison of observed BSS luminosity and temperature
distributions with theoretical models reveals a population of
luminous, hot BSS which is not easily interpreted.

\end{abstract}


\section{Introduction} 
\label{intro}
Blue straggler stars (BSS) were first detected in the Galactic Globular Cluster
(GGC) M3 \citep{sandage53} as a sparsely populated sequence extending to higher
luminosity than the turnoff (TO) point of normal hydrogen--burning main
sequence stars in the Color--Magnitude diagram (CMD). Therefore their position
in the CMD suggests that they are massive stars still burning hydrogen in their
core in an old star cluster. Since no other evidence of subsequent star
formation episodes into the GGCs stellar populations can be found, one of the
great challenges of the last 50 years was to understand the origin of BSS.

The first surveys for BSS, done with photographic plates, were limited
to the outer parts of the clusters and no tendency for the BSS to
concentrate toward the central regions was observed. The advent of
telescopes of higher resolution and in particular the launch of the
Hubble Space Telescope (HST) allowed the inspection of the cores of
GGCs, revealing that BSS are more centrally concentrated than the
normal stars, e.g., the subgiants, of the same luminosity.  This
suggests that these stars are generally more massive than the cluster
subgiants, leading many authors to consider BSS as the offspring of
binary systems. Two mechanisms for making BSS have been suggested: (i)
mass transfer between or the merger of two stars in a primordial
binary (where``primordial'' refers to binaries created when the
cluster formed) and (ii) collisions in regions of very high stellar
density \citep{hills76, fusipecci92, ferraro93, ferraro95, bailyn95,
meylan97}. The class of collisional BSS can be further subdivided into
those produced by direct collisions, those created as collisions
harden primordial binaries until they merge, and those resulting when
binaries are produced in a collision and merge later.

The study of BSS in stellar clusters provides new insights not only
into dynamical interaction and evolution of individual stars, but also
of the cluster as a whole. Indeed gravitational interactions between
cluster stars force the GGCs to evolve dynamically on timescales
generally smaller than their ages. The first evidence of dynamical
processes within a GGC is the segregation toward the center of the
more massive stars (or binaries) \citep[see][ and references therein,
for a review]{bailyn95}. A star cluster can undergo other dynamic
evolution: galactic tidal stripping continuously removes stars from
the outer region of the cluster, and other stars are lost because their
velocities are higher than the escape velocity of the cluster. As a
consequence the cluster is forced to adjust its structure and the core
must contract. In some cases this process can lead to a catastrophic
collapse of the core. Binaries are thought to play a fundamental role
in the core collapse---binary-binary collisions could be effective in
delaying the collapse of the core, avoiding infinite central
density. In this case, while the core tries to collapse, most of the
binaries in the central regions will be destroyed by close encounters,
and the survivors will become tightly bound, producing an
overabundance of BSS \citep[i.e. the case of M80,][]{ferraro99b}.

The GGC NGC~6752 is a very interesting target to study the role of
dynamical effects on stellar evolution. The cluster hosts many
faint X-ray sources in the core, which may be cataclysmic variables
\citep{pooley02}. In addition there are at least five millisecond
pulsars (MSPs), two of which are in the outer part of the
cluster. These two are the most radially distant MSPs
gravitationally bound to a GC. Moreover, two of the three central MSPs
display an anomalous acceleration \citep{damico02}, which suggests a
very high cluster central mass-to-light ratio
\citep{colpi02}. \citet{ferraro03a} showed that NGC~6752 is a
dynamically evolved cluster, probably undergoing a post-core-collapse
bounce. They investigated scenarios for simultaneously explaining both
anomalous acceleration of the inner MSPs and the ejection of the most
external MSPs, concluding that the existence of a binary black hole of
intermediate mass could be a viable possibility \citep[see
also][]{colpi02, colpi03}.

Since BSS are excellent tools to investigate the dynamical status of a
cluster, in this paper we will focus our attention on this
population. As done in previous work \citep{ferraro93, ferraro97,
ferraro03c}, we have combined high resolution HST data with wide field
images in order to study the behavior of populations over the entire
cluster extent. The data are presented in \S \ref{obs}, while in \S
\ref{sample} the BSS candidates and the criteria of selection are
shown. The radial distribution of the BSS is presented in \S
\ref{bssrad}, while we have compared our results with those of other
clusters (\S \ref{othergc}). The results of the collisional models are
shown in \S \ref{model}.  These results are discussed in \S
\ref{discussion}.
 
\section{Observations and data analysis}
\label{obs}

To search for BSS in NGC~6752 we have used two data sets: 

{\it (i) High resolution set---} a series of high-resolution
WFPC2-HST images were obtained on March 2001, using the F555W
($V$), F336W ($U$) and F255W ($mid-UV$) filters as part of a long term
project (GO-8709, PI: F. R. Ferraro) aimed at studying the central
stellar populations in a set of GGCs. In this data set the planetary
camera (PC, which has the highest resolution $\sim 0\farcs{046}/{\rm
pixel}$) was roughly centered on the cluster center while the Wide
Field (WF) cameras (at lower resolution $\sim 0\farcs{1}/{\rm pixel}$)
sampled the surrounding outer regions;

\begin{figure}
\plotone{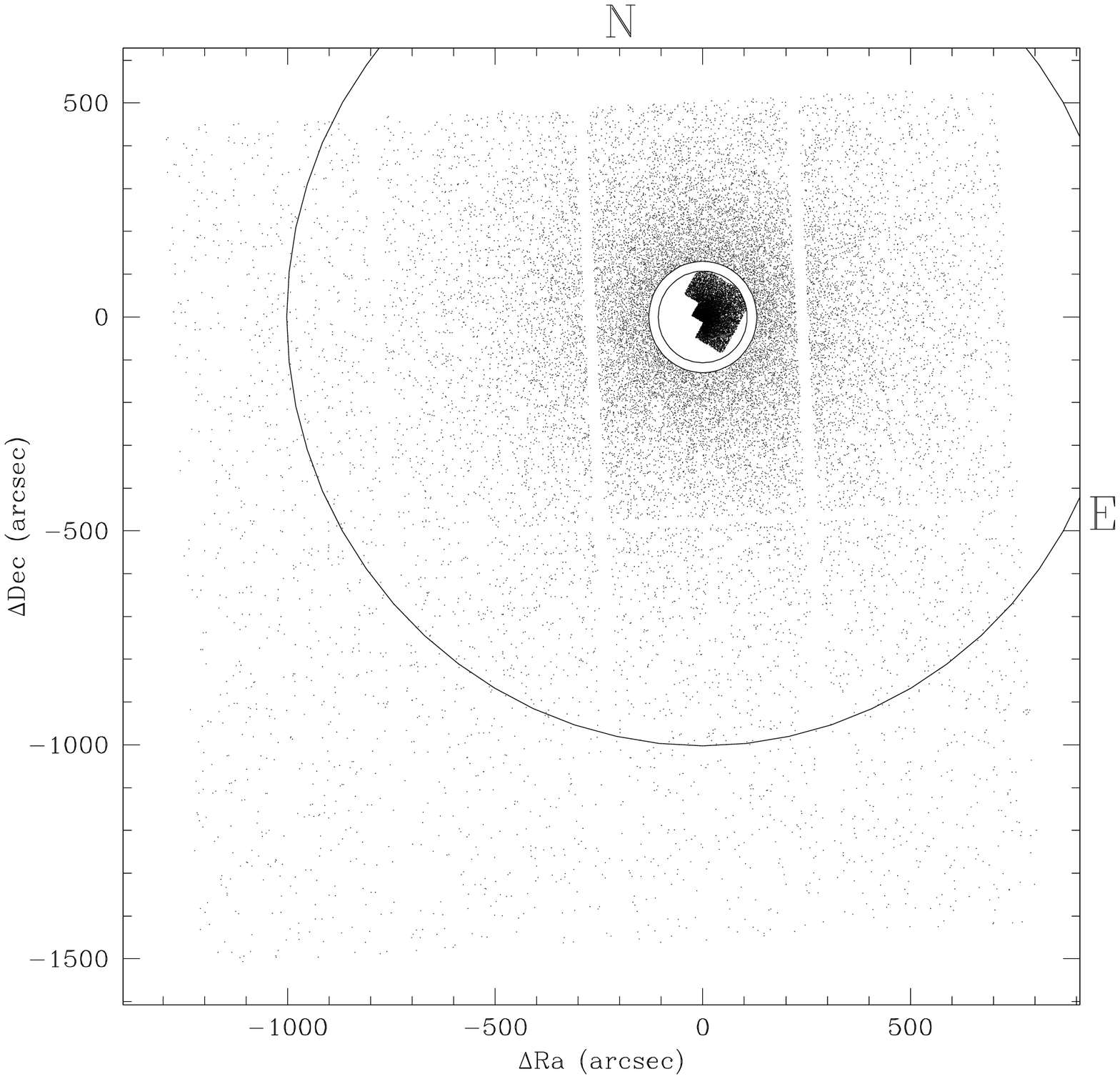}
\caption{\label{map} Computed map of the HST and WFI FoV. The black circular
line has a radius of $16\farcm67$ and it is centered on the cluster $C_{\rm
grav}$ determined by \citet{ferraro03a}}

\end{figure}

{\it (ii) Wide Field set---} a complementary set of multi-filter ($B$, $V$,
$I$) wide field images was secured during an observing run at the 2.2m ESO-MPI
telescope at ESO (La Silla) in July 1999, using the Wide Field Imager (WFI). 
The WFI is a mosaic of 8 CCD chips (each with a field of view of $8'\times
16'$) giving a global field of view of $33'\times 34'$. The cluster was roughly
centered on chip $\#2$ (see Figure~\ref{map}). 

Standard IRAF\footnote{IRAF is distributed by the National Optical Astronomy
Observatory, which is operated by the Association of Universities for Research
in Astronomy, Inc., under cooperative agreement with the National Science
Foundation.} tools were used to correct the raw WFI images for bias and flat
field and for trimming the over-scan region. The photometric reduction of the
images has been performed with the DAOPHOT package in the IRAF environment,
applying the point spread function (PSF) fitting procedure independently on
each $V$ and $B$ images. For each chip the objects found in both bands were
cross-identified with a package developed at the Bologna Observatory
(Montegriffo et al. 2004, in preparation). After this match a catalog listing
the instrumental $B,~V$ magnitudes for all the stars in each field has been
obtained. The WFI catalog was finally calibrated by using the data set of
\citet{buonanno86}.

The photometric reductions of the high resolution images were carried
out using ROMAFOT \citep{buonanno83}, a package developed to perform
accurate photometry in crowded fields and specifically optimized to
handle under-sampled PSFs \citep{buonanno89} as in the case of the
HST-WF chips.

PSF-fitting instrumental magnitudes have been obtained using the  standard
procedure described in \citet{ferraro97, ferraro01}.  The final catalog of the
F555W, F336W and F255W magnitudes was calibrated by using the zero-points 
listed by \citet{holtzmann95}. 

The {\it Guide Star Catalog} ($GSCII$) was used to search for astrometric 
standards in the entire WFI image field of view. The procedure used to
obtain the astrometric solution of the 8 WFI chips is completely described in
\citet{ferraro03a}. 

Since the small field ($2\farcm5$ on the side) of the high resolution WFPC2/HST
images was entirely contained within the field of view of the WFI chip $\#2$
(see Figure~\ref{map}), we used more than 1200 bright stars in the WFI catalog
lying in the WFPC2-FoV as {\it secondary astrometric standards} in order to
properly find an astrometric solution for the WFPC2 catalog. The estimated
global uncertainties in the astrometric procedure are less than $\sim 0\farcs4$
both in RA and Dec.  This procedure allowed us to obtain two catalogs (WFPC2
and WFI) which are fully homogeneous in the absolute coordinate system.

Results from the analysis of the F555W and F336W catalog were presented in a
previous paper \citep{ferraro03a}. Here, since our goal is the
identification and the analysis of the BSS population, we will focus on the
($m_{255}$, $m_{255} - m_{336}$) plane.

\section{The CMD and the BSS candidates}
\label{sample}

\subsection{The HST data} 
\begin{figure}
\plotone{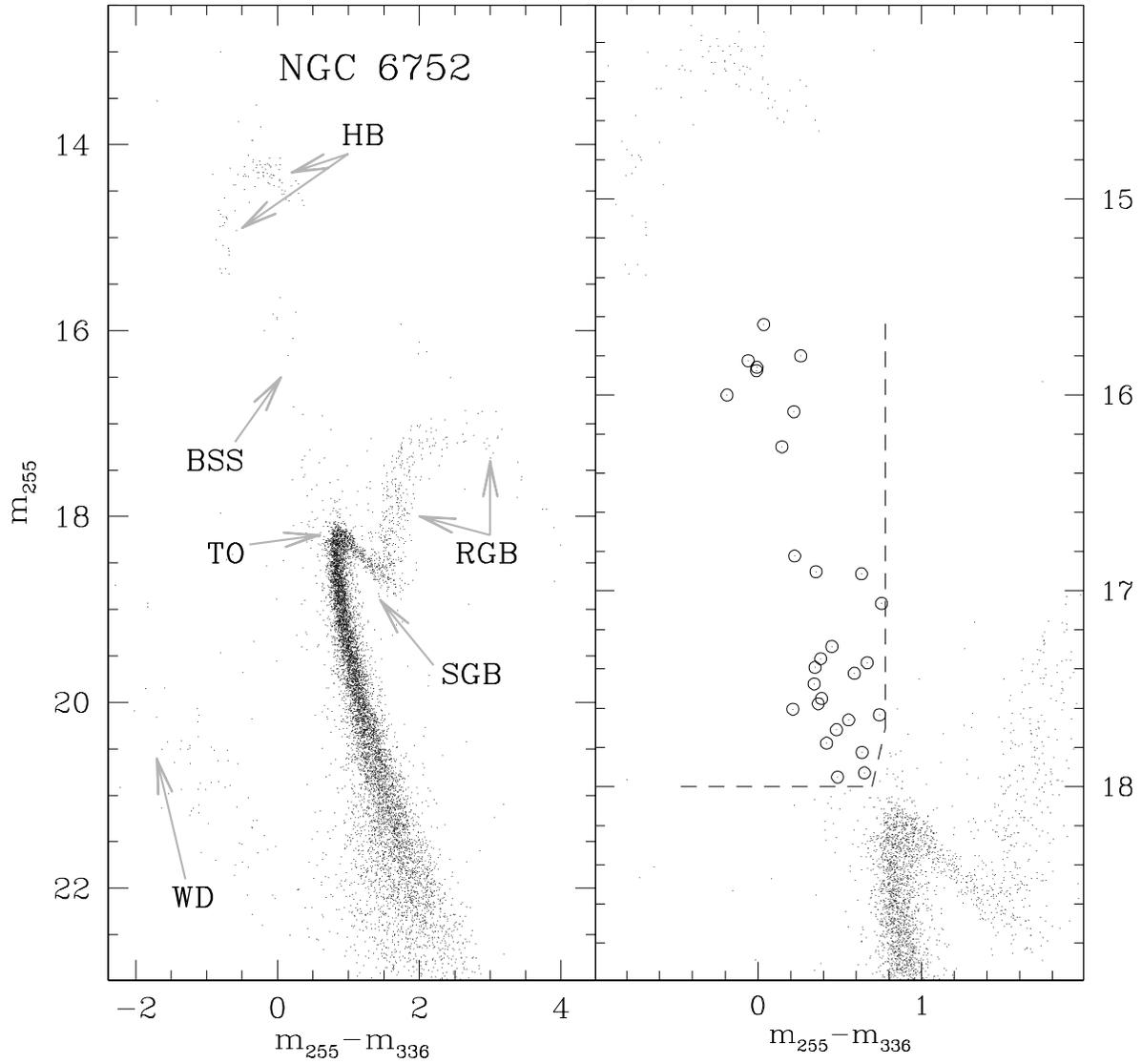} 
\caption{\label{CMD}(m$_{255}$, m$_{255}$-m$_{336}$) CMD for the central region
of NGC~6752, from WFPC2/HST observations. {\it Left panel:} The whole CMD. The
main branches are indicated. {\it Right panel:} The zoomed CMD in the BSS
region. The selected BSS are marked with large empty circles.}

\end{figure}

Many studies \citep[][ and references therein]{dorman95} have shown
that the main contributors to the UV emission of GCs are the hot stars
which populate the horizontal branch (HB) and the BSS \citep[see for
example Figure~1 by][]{ferraro99}. In UV planes the main branches
display very different morphology from the usual optical CMD
(i.e. $V,~ V-I$).  Figure~\ref{CMD} shows the UV-CMD of NGC~6752 in
the ($m_{255}$, $m_{255} - m_{336}$) plane for more than 11,000 stars
identified in the HST field of view. As can be seen, the red giant
branch (RGB) is very faint, while the horizontal branch (HB),
excluding the hottest section, which bends downward because of the
increasing bolometric correction, appears diagonal. Since red giants
are faint in UV, the photometric blends, which mimic BSS in visible
CMDs, are less problematic. Thus, a complete BSS
sample can be obtained even in the densest cores. Indeed, the
($m_{255}$, $m_{255} - m_{336}$) plane is an ideal tool for selecting
BSS \citep{ferraro97, ferraro99, ferraro03c}. 

The BSS candidates occupy a narrow, nearly vertical, sequence spanning $\sim 3$
mag in $m_{255}$. Two limits (one in color, and one in magnitude) have been
assumed to properly select the BSS sample in the UV-CMD shown in
Figure~\ref{CMD}.  The BSS sequence blends smoothly into the main sequence (MS)
near the cluster TO. To select only `safe' BSS, we have chosen only stars
brighter than $m_{255} \sim 18$ (0.3 mag brighter than the cluster TO). This
limiting magnitude is consistent with that used in our other BSS catalogs
\citep{ferraro97, ferraro03c} so as to allow easy comparison. To exclude
spurious detections due to blended sources, we have chosen only objects bluer
than $m_{255} - m_{336} \sim 0.76$. In addition we have excluded two objects
which satisfy these criteria but which lie near the boundary of the region and
are in regions of severe crowding.

Following these criteria we have identified 28 BSS in the HST-WFPC2
FoV (hereafter HST sample). The BSS candidates are listed in
Table~\ref{tab1}: the first two columns list the identification
numbers, while columns (3)--(5) report the $m_{255}$, $m_{336}$ and
$m_{555}$ magnitudes respectively. In columns (6) and (7) we report
the astrometric coordinates (Right Ascension and Declination).

\subsection{The WFI data}

In order to avoid spurious effects due to incompleteness of the ground based
observations in the crowded central region of the cluster, we restricted the
WFI sample to stars with $r > 130\arcsec$ from the center of gravity ($C_{\rm
grav}$) given in \citet{ferraro03a} (see Figure~\ref{map}). Since the outer
regions of the WFI FoV are dominated by field stars, in this paper we consider
as cluster members only stars contained within the tidal radius $r \sim
16\farcm67$ ($\sim 1000\arcsec$) derived from the radial density profile by
\citet{ferraro03a}.

\begin{figure}
\plotone{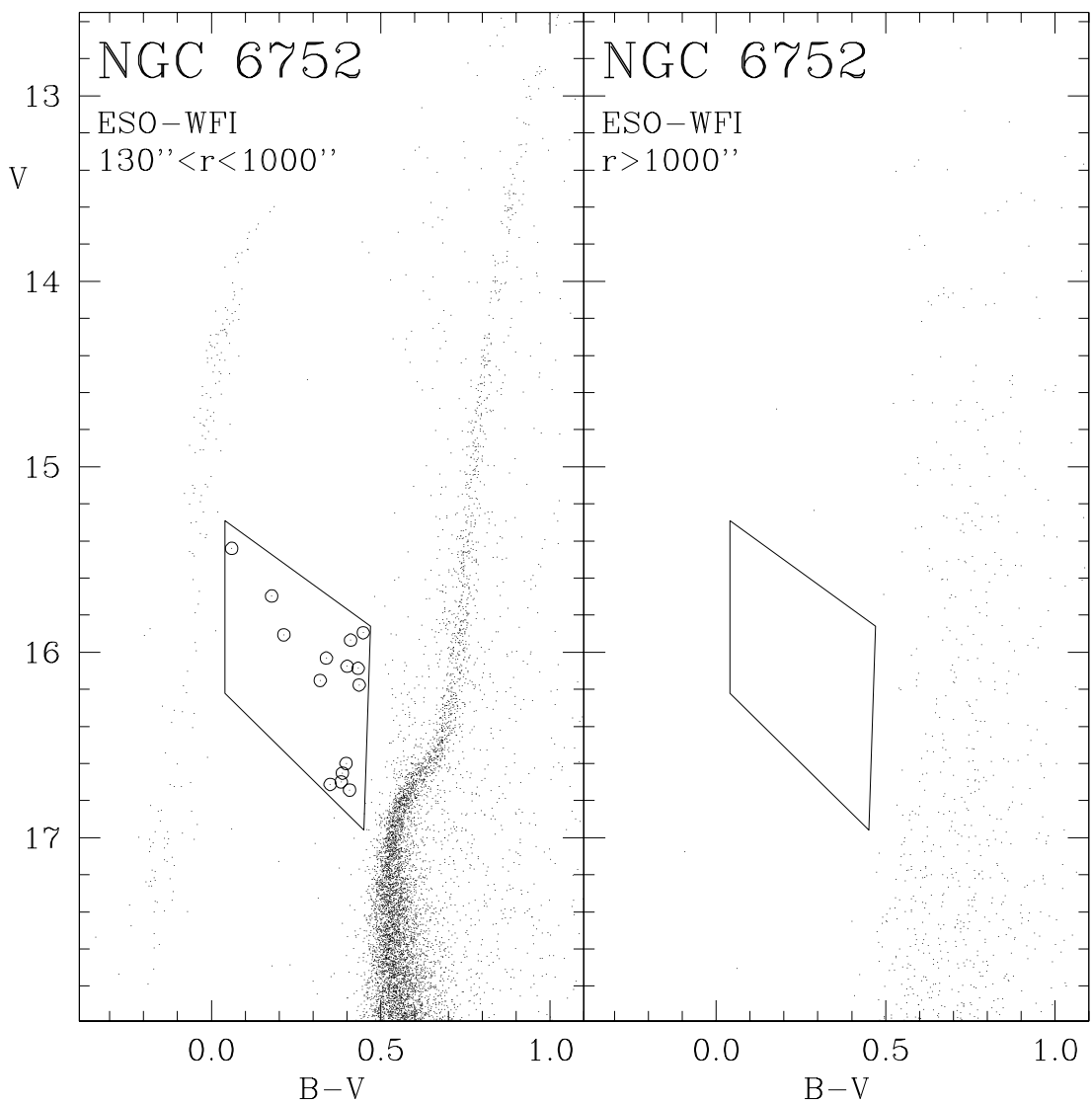} 
\caption{\label{wficmd}($V,~B-V$) CMD for the external region of NGC~6752 from
ground based (WFI) observations. {\it Left panel}: All the stars with
$130\arcsec < r < 16\farcm67$ from the cluster center $C_{grav}$ have been
plotted. The selection box for the BSS is also shown. Selected BSS are marked
with large empty circles. {\it Right panel}: Stars with $r > 16\farcm67$ are
plotted. As can be seen no stars lie in the BSS selection box, suggesting that
our BSS sample is not significantly contaminated by field stars.}

\end{figure}

BSS in the WFI sample have been selected from the ($V, B - V$) CMD
using the selection box shown in Figure~\ref{wficmd}---{\it left
panel}.  The sample includes only stars brighter than $V \sim 16.9$,
and bluer than $(B-V)\sim 0.44$. There are only 15 BSS candidates. No
stars are found in this region of the CMD for $r \ga 16\farcm67$ (see
Figure~\ref{wficmd}---{\it right panel}) suggesting that field
contamination should not significantly affect the selected BSS sample.

The BSS candidates detected in the WFI-FoV are listed in Table~\ref{tab2}. As in
Table 1 the first two columns list the identification numbers,  $B$ and $V$
magnitudes are reported in columns (3) and (4) respectively. The last two
columns ((5) and (6)) list the astrometric coordinates.

\section{BSS radial distribution}
\label{bssrad}

\begin{figure}
\plotone{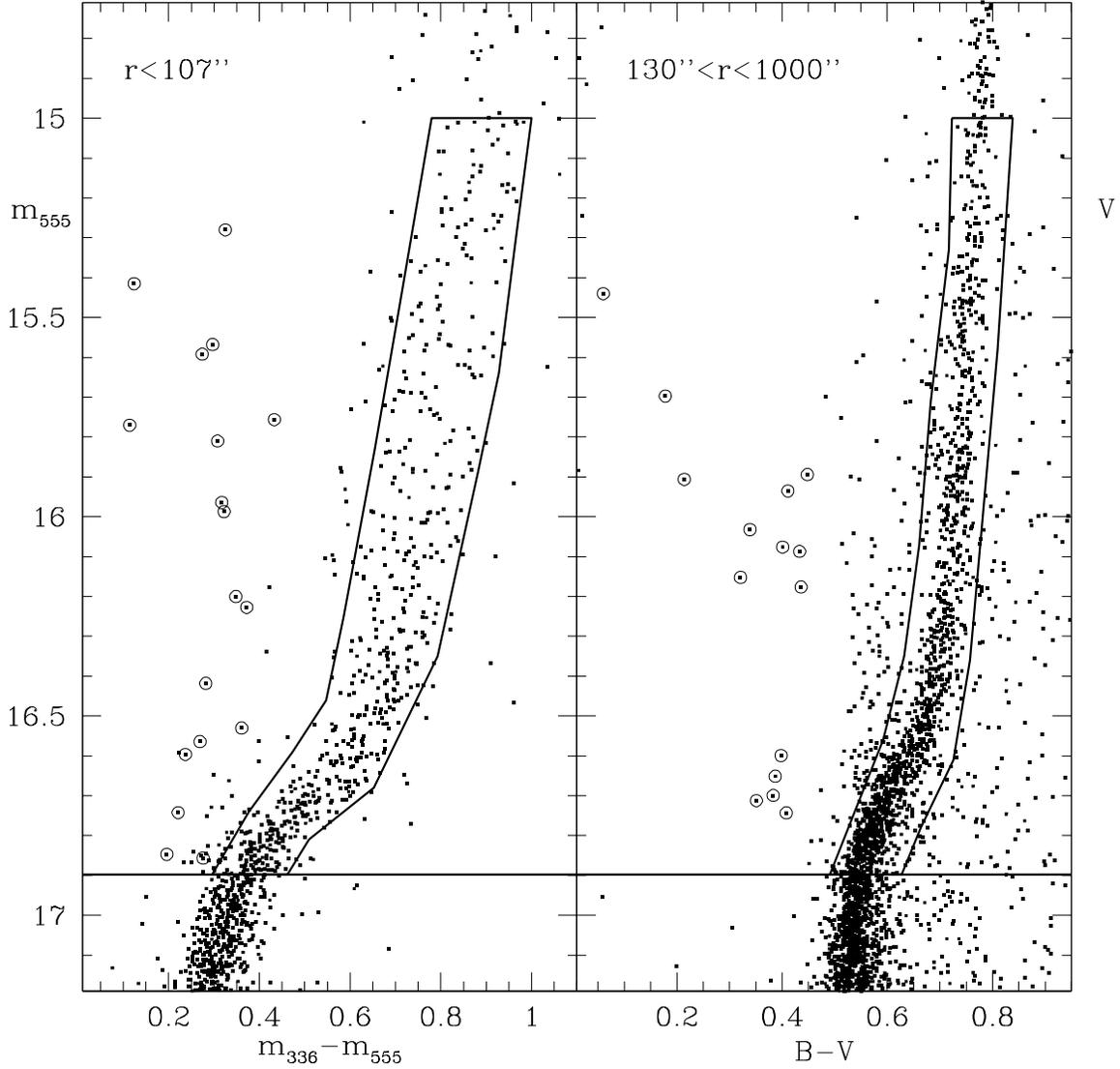} 
\caption{\label{hst+wfi}Direct comparison between the HST ({\it left panel})
and the ground based WFI ({\it right panel}) samples. Selected BSS populations
are marked with large empty circles. Note that the BSS selection for the
central region has been done in the UV plane - see Figure~\ref{CMD}. Selection
boxes for the RGB reference stars are also shown. The horizontal line marks the
BSS threshold.}

\end{figure}

In order to study the BSS radial distribution in NGC~6752 over the entire
cluster extension, we must combine the two samples (HST $+$ WFI). This requires
that the two BSS samples have the same limiting magnitude. To do this we can
use the $V$ band which is in common to the two data sets (since the F555W-WFPC2
filter is approximately a $V$ filter).  Figure~\ref{hst+wfi} shows the two CMDs
in the ($m_{555}$, $m_{336} - m_{555}$), and in the ($V$, $B - V$) planes for
the HST and WFI samples respectively. As can be seen the faint threshold for
the ground based BSS sample roughly corresponds to $V \sim 16.9$. For the most
part in the following, we consider only the HST-BSS brighter than this limit
(19 BSS from BSS-HST-1 to BSS-HST-19 in Table~1). Note that only 18 HST-BSS are
plotted in Figure~\ref{hst+wfi}, because the bright BSS-HST-3 was not measured
in the F555W filter since it lies near a bright red star which seriously
affects its $V$ magnitude. (However BSS-HST-3 is quite bright in the UV and it
certanly has $V<16.9$. Note that the same happen to BSS-HST-22 which is
significantly fainter than BSS-HST-3 and by comparing HST-BSS-22 with stars of
similar UV properties and measurable $V$, we estimate that it lies just on the
faint side of the $V<16.9$ boundary, and we eliminate it from our sample). The
final selection yeld a total catalog of 34 BSS (19 from HST and 15 from the
WFIsample)

Fig.~4 illustrates another important point---the danger of using
optical CMDs to identify BSS in crowded regions. Even using HST, in
dense GGC cores the BSS region can be populated by spurious objects
due to blending of SGB, RGB and HB stars.  This is quite clearly shown
by the comparison of Fig.~2 and the left panel of Fig.~4. In Fig. 4
there are many objects lying in the region of the BSS which are not
circled. Only the BSS selected in the UV diagram are circled and
genuine. All of the other objects (indeed the majority of objects
fainter than $V \sim 16.2$) are blends. 

To study the BSS radial distribution we must describe the BSS relative
to some reference stellar populations. Here we use as ``reference''
both HB and RGB stars. We decided to use both populations since the HB
is clearly defined and it has been used in previous papers
\citep[see][]{ferraro03b}; on the other hand the RGB, including the
lower RGB, is much more populous than HB, hence star counts are less
affected by statistical fluctuations. In the HST catalog the RGB stars
have been selected in the ($m_{555}$, $m_{336} - m_{555}$ plane--see
Figure~\ref{hst+wfi}) to reduce any bias which might be
introduced by the poor photometry for the redder stars in the UV
bands. In matching the two samples the lower boundary of the selection
box is most important, and Figure~\ref{hst+wfi} shows that it is well
matched in the two samples. The selection in colors is somewhat
arbitrary, but in both the cases the selections include the bulk of
the RGB population. A few stars were excluded; they could be either
poorly measured stars and/or field stars, especially in the WFI
sample. The fact that RGB radial distribution is the same as the HB
distribution (see below) suggests that there is no significative bias
between the two samples in our selection.  

The resulting comparison population samples are 87 HB and 255 RGB in
HST and 264 HB and 1984 RGB in WFI respectively.

\begin{figure} 
\plotone{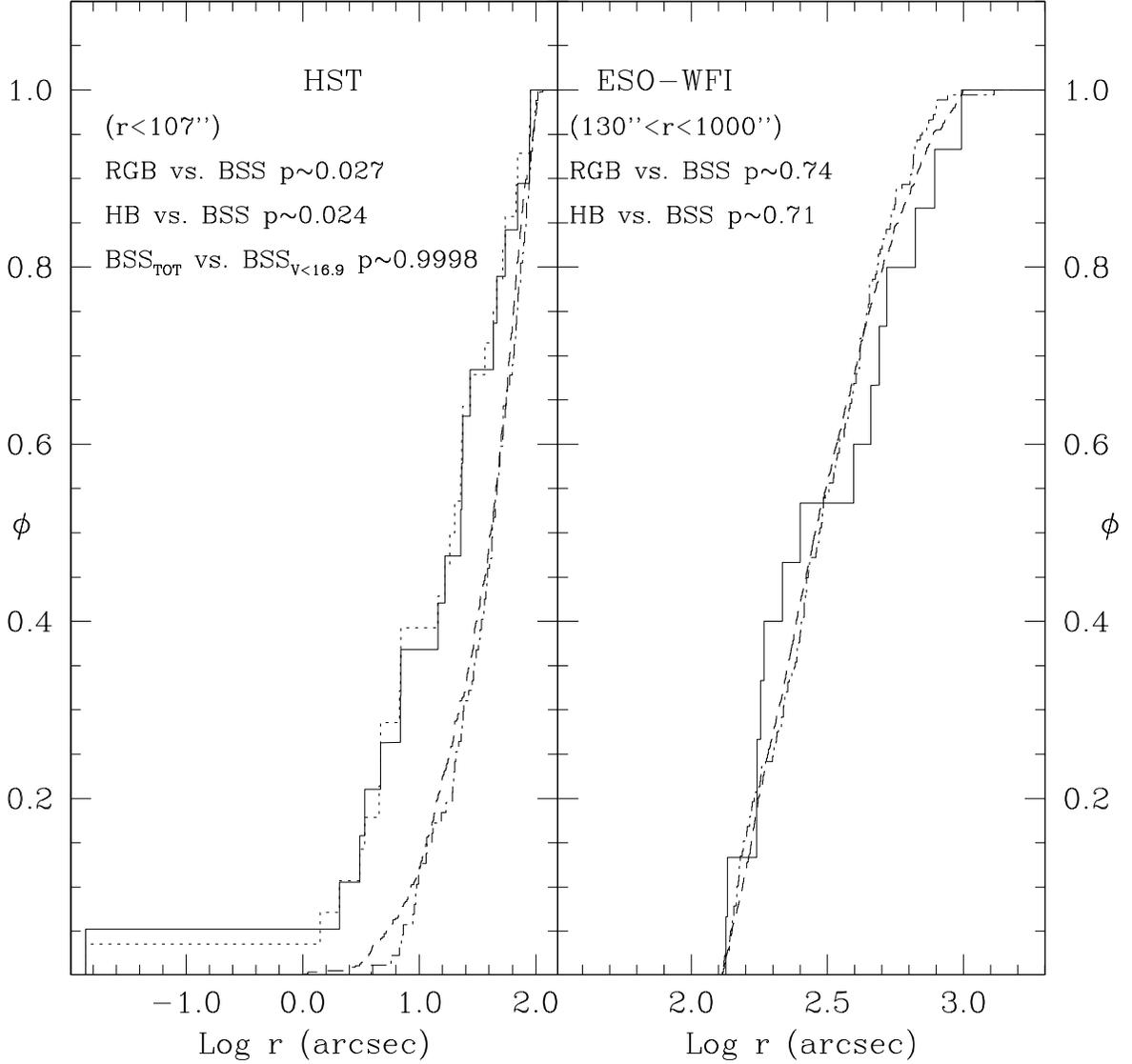}  
\caption{\label{rad}Cumulative radial distributions for the final BSS
({\it solid lines}) sample with respect to RGB stars ({\it dashed
line}) and to HB stars ({\it dot-dashed line}) as a function of their
projected distance ($r$) from the cluster C$_{\rm grav}$. In the {\it left
panel} we also show the radial distribution of the total HST-BSS sample (i.e.
all the 28 BSS listed in Table~1) ({\it short-dashed line}). $p$
is the probability that two samples are extracted from the same parent
population.}

\end{figure}

In Figure~\ref{rad} we have plotted the cumulative radial
distributions both for the BSS and reference populations for the HST
($r < 107\arcsec$---{\it left panel}) and for the WFI ($130\arcsec < r
< 16\farcm17$---{\it right panel}) samples. The radial distributions
were computed by adopting the cluster $C_{\rm grav}$ recently
determined by \citet{ferraro03a}. In central regions we show both the
full 28 star HST BSS distribution (short-dashed line) and the 19 star
truncated BSS-distribution ($V < 16.9$ --- solid line). These two
distributions are essentially identical. A Kolmogorov-Smirnov (K-S)
test shows that the two samples could have been drawn from the same
distribution $\sim 99.98$\% of the time.  This shows that the exact
value of the limiting magnitude chosen does not affect our results.
Both the full and truncated HST samples are more concentrated than the
RGB (dashed line) and the HB (dot-dash line) stars selected in the
same area. K-S tests yield probabilities of $\sim 97.3$\% and $\sim
97.6$\% that the truncated-BSS sample in the central region of
NGC~6752 has a different radial distribution with respect to the
reference HB and RGB population respectively. As noted earlier (see
Figure~1) the shape of WFPC2 prevents us from fully sampling the
smallest annuli. In the region between $30\arcsec$ and $130\arcsec$ we
sample only 40\% of the total area including a 23\arcsec\ wide
transition area between the HST and WFI data which we do not sample at
all. Because the number of BSS stars is small (8 between $30\arcsec$
and $130\arcsec$) the counting errors can be large, and if the numbers
are corrected for undersampling these errors are magnified. However,
our conclusions are drawn from a comparison with a reference
population, and both the HB and RGB reference populations are drawn
from exactly the same areas as the BSS. Such comparisons are not
affected by the under/un-sampled annuli.

On the other hand, in the outer sample (right
panel) the radial distribution of the BSS is not statistically
different (less than 1$\sigma$) from those of either the RGB or HB. We
note that this effect could be also due to the fact that the number of
BSS is too small to draw a definite conclusion.


In two previously surveyed clusters (M3 and 47~Tucanae) we have shown that the
BSS are more concentrated than the reference population in the central regions.
The reverse is true in the outer regions with the BSS being less concentrated
than the reference population.  The situation is less well defined in
NGC~6752.  The comparison of the radial distribution shown in Figure~\ref{rad}
clearly demonstrates ({\it left-panel}) that BSS are much more concentrated
toward the center with respect to the normal cluster stars. The small sample in
the outer region (Figure~\ref{rad}-{\it right panel}) shows no evidence that
the BSS are less concentrated than the reference populations, but an effect
like that in M3 and 47~Tuc could well not show up in a sample of this size.

\begin{figure}
\plotone{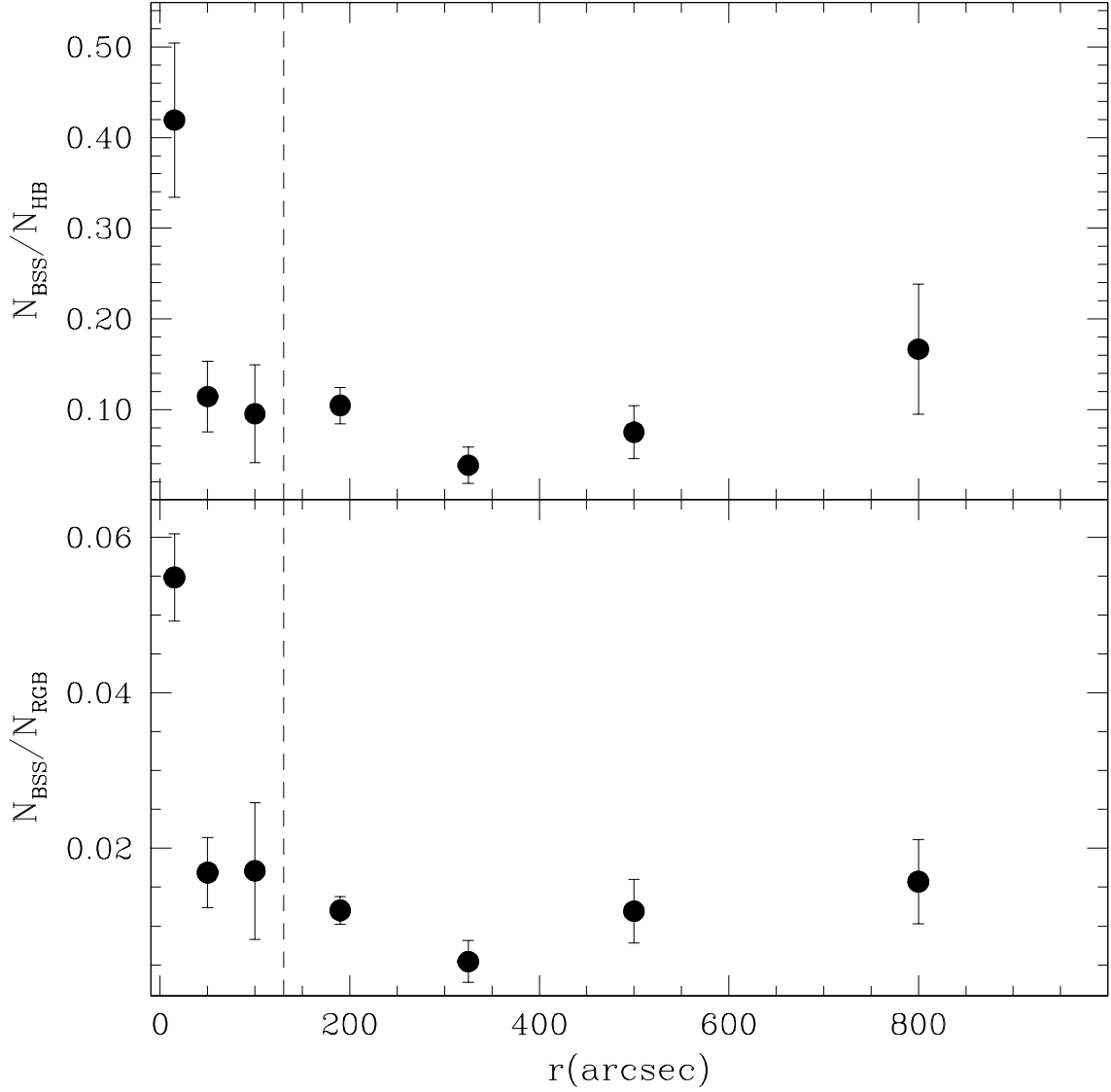}
\caption{\label{ratio1}Relative frequency of the BSS with respect to HB ({\it 
upper panel}) and RGB stars ({\it lower panel}) plotted as a function of the 
distance from the cluster center. The vertical dashed lines
distinguish the cluster region  observed with HST (by using
UV filters) from the region observed from the ground (by
using optical B,V filters).}

\end{figure}

In order to further investigate the distribution of BSS, we computed
the radial behavior of the BSS relative frequency $F^{\rm BSS}_{\rm
HB}={\frac{N_{\rm BSS}}{N_{\rm HB}}}$, where $N_{\rm BSS}$ and $N_{\rm
HB}$ is the number of BSS and HB stars respectively. In doing this we
have subdivided the surveyed area into a set of concentric annuli
(each containing roughly $\sim 10$\% of the reference
populations), and we have counted the number of BSS and HB stars
contained in each annulus. The relative frequency as function of
distance is shown in Figure~\ref{ratio1}---{\it upper panel}. The
distribution shows a bimodal trend---it reaches the maximum in the
innermost annulus ($N_{\rm BSS}/N_{\rm HB} \sim 0.42$) and quickly decreases
to less then 0.04 as $r$ increases. Nevertheless, in the most external
annuli the distribution shows a small upturn reaching $\sim 0.2$. Errors 
in the relative frequency of BSS with the respect of HB, here written as $R=a/b$
(where $a$ indicate the BSS and $b$ the HB numbers respectively) are
$$
\sigma_R = (R^2\sigma^2_b+\sigma^2_a)^{1/2}/b
$$

In order to decrease the statistical fluctuation in the distribution
due to the small numbers, and to show that the observed bimodality is
not introduced by an anomalous HB star distribution, we also have
normalized the BSS number to the RGB population selected as shown in
Figure~\ref{hst+wfi}. The result is shown in Figure
\ref{ratio1}---{\it lower panel} (errors were derived adopting the same
formula of the previous case). As can be seen the bimodal behaviour
of the BSS frequency is fully confirmed independent of the
reference population. This result (though of lower significance)
closely follows the bimodality observed in M3, M55 and recently in
47~Tuc, thus increasing the number of clusters showing this peculiar
radial behaviour. Extensive surveys in the outer region of other
clusters (as M80, M5, etc.) are needed before we can conclude that
this is the ``natural'' radial distribution of BSS in globulars.

\section{Comparison with other GCs}
\label{othergc}

In a recent paper \citet{ferraro03b} have compared the BSS populations in the
central region of 6 GGCs with different central density and metallicity. Here
we compare the BSS population detected in the HST FoV for NGC~6752 with those
presented by Ferraro et al. for the other clusters. We also include 47~Tuc
\citep{ferraro03c} in the discussion with the caveat that it was observed in a
different photometric plane which might introduce some additional uncertainty.

\begin{figure}
\plotone{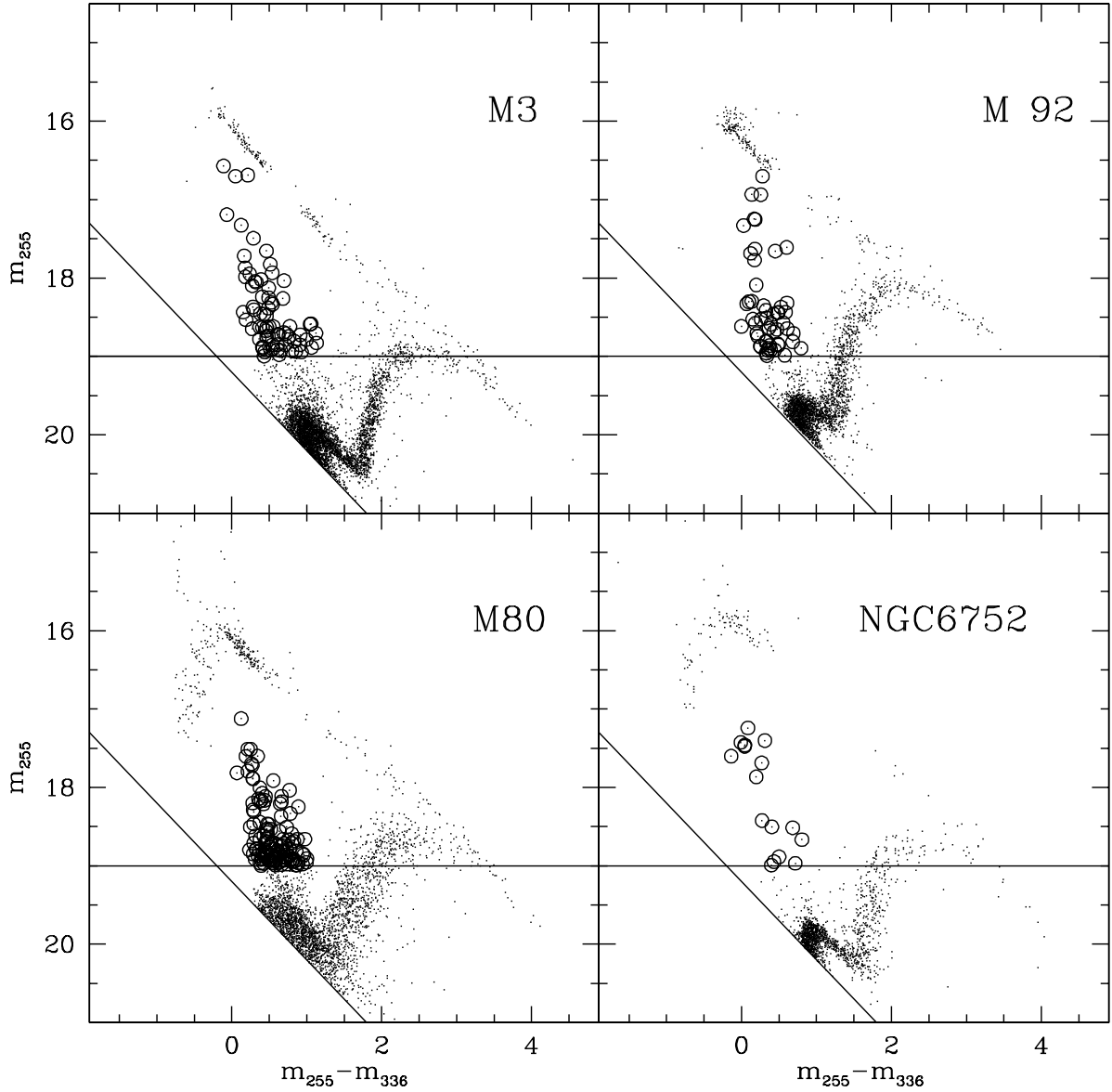} 
\caption{\label{m3}(m$_{255}$, m$_{255}$-m$_{336}$) CMDs for four high density
GGCs, namely M3, M92, M80, and NGC~6752. Horizontal and vertical shifts have
been applied to the CMDs in order to match the sequences of M3. The horizontal
solid line corresponds to m$_{255}$=19 in M3. The BSS brighter than this value
are marked as large empty circles.}
\end{figure}

To do this we have applied the same criteria adopted by Ferraro et
al.: the HST-CMD of NGC~6752 was shifted in $m_{255}$ magnitude and
color to match the MS of M3 (see Figure~\ref{m3}). In this
comparison only bright BSS (bBSS) with magnitude $m_{255}$ brighter
than 19.0 are considered. This selection reduced our sample to 16
stars. (Note that the faint and bright subsamples show radial
distributions which are almost identical, suggesting that our BSS
sample is not significantly contaminated with MS stars.)

In Table~\ref{comparison} we compare some of the properties of the
previously observed clusters and their BSS with NGC~6752. We give
central density, $\log\,\rho_0$, cluster mass, central velocity
dispersion, and $\sigma_0$, from \citet{pryor93}. $F_{\rm HB}^{\rm
bBSS}$ as defined above is determined for the entire HST sample and is
not the central value.

In the case of NGC~6752 the ratio $F^{\rm bBSS}_{\rm HB}$ is
equal to 0.18, one of the lowest ever derived for a cluster. Of the
the six clusters analyzed by \citet{ferraro03b}, only M13 has a
lower $F^{\rm bBSS}_{\rm HB}$ value. 

An other useful parameter to compare the BSS population in GGCs is the radius
containing half the BSS sample ($r^{\rm bBSS}_{1/2}$). For NGC~6752 $r^{\rm
bBSS}_{1/2}\sim 22\farcs 7$. If fit by one single mass King Model the core
radius $r_c$ of NGC~6752 is $13\farcs 7$ \citep{ferraro03a} and the ratio
$r^{\rm bBSS}_{1/2}/r_c \sim 1.67$. This is the largest value ever measured for
this ratio \citep{ferraro03b}. However Ferraro et al. (2003b) noted that one
King model did not fit the NGC~6752 profile well. A much better fit was
achieved using two single mass King models, one to fit the core, the other to
the external region. This unusual profile suggests an unusual dynamical state,
perhaps a core bounce.  Could this be related to the large value of $r^{\rm
bBSS}_{1/2}/r_c$? Indeed if one uses the larger of the double fit core radii
($r_c \sim 28 \arcsec$, see Table~\ref{comparison}) the ratio $r^{\rm
bBSS}_{1/2}/r_c$ turns out to be $\sim 0.82$, fully comparable to the other
clusters.

\section{Collisional Models}
\label{model}

We have compared the observed BSS population in NGC 6752 with models
of BSS populations. The models used here are described in
detail in \citet{sills99} and have been applied to 47~Tuc
\citep{sills00} and six other clusters (M3, M10, M13, M80, M92 and NGC 288) 
\citep{ferraro03b}.  We assume that all the BSS were formed 
via direct stellar collisions between two stars during an encounter
between a single star and a binary system.  The trajectories of the
stars during the collision are modeled using the STARLAB software
package \citep{mcmillan96}. The masses of the stars involved are chosen
randomly from a mass function for the current cluster and a different
mass function which governs the mass distribution within the binary
system.  A binary fraction, and a distribution of semi-major axes must
also be assumed.  The output of these simulations is the probability
that a collision between stars of specific masses will occur. We have
chosen standard values for the mass functions and binary
distribution. The current mass function has an index $x=-2$, and the
mass distribution within the binary systems are drawn from a Salpeter
mass function ($x=1.35$). We chose a binary fraction of 20\% and a
binary period distribution which is flat in $\log\, P$. The total stellar
density was taken from the central density of each cluster. The effect
of changing these values is explored in \cite{sills99}.  The collision
products are modeled by entropy ordering of gas from colliding stars
\citep{sl97} and evolved from these initial conditions using the Yale
stellar evolution code YREC \citep{guenther92}. The models reported here used
a metallicity of ${\rm [Fe/H]}=-1.56$ \citep{harris96}.  By weighting the resulting
evolutionary tracks by the probability that the specific collision
will occur, we obtain a predicted distribution of BSS
In order to explore the effects of non-constant BSS
formation rates, we examined a series of truncated rates.  In these
models we assumed that the BSS formation rate was constant
for some portion of the cluster lifetime, and zero otherwise.  This
assumption is obviously unphysical---the relevant encounter rates
would presumably change smoothly on timescales comparable to the
relaxation time.  However these models do demonstrate how the
distribution of BSS in the CMD depend
on when the BSS were created, and thus provide a basis for
understanding more complicated and realistic formation rates.

\begin{figure}
\epsscale{.85}
\plotone{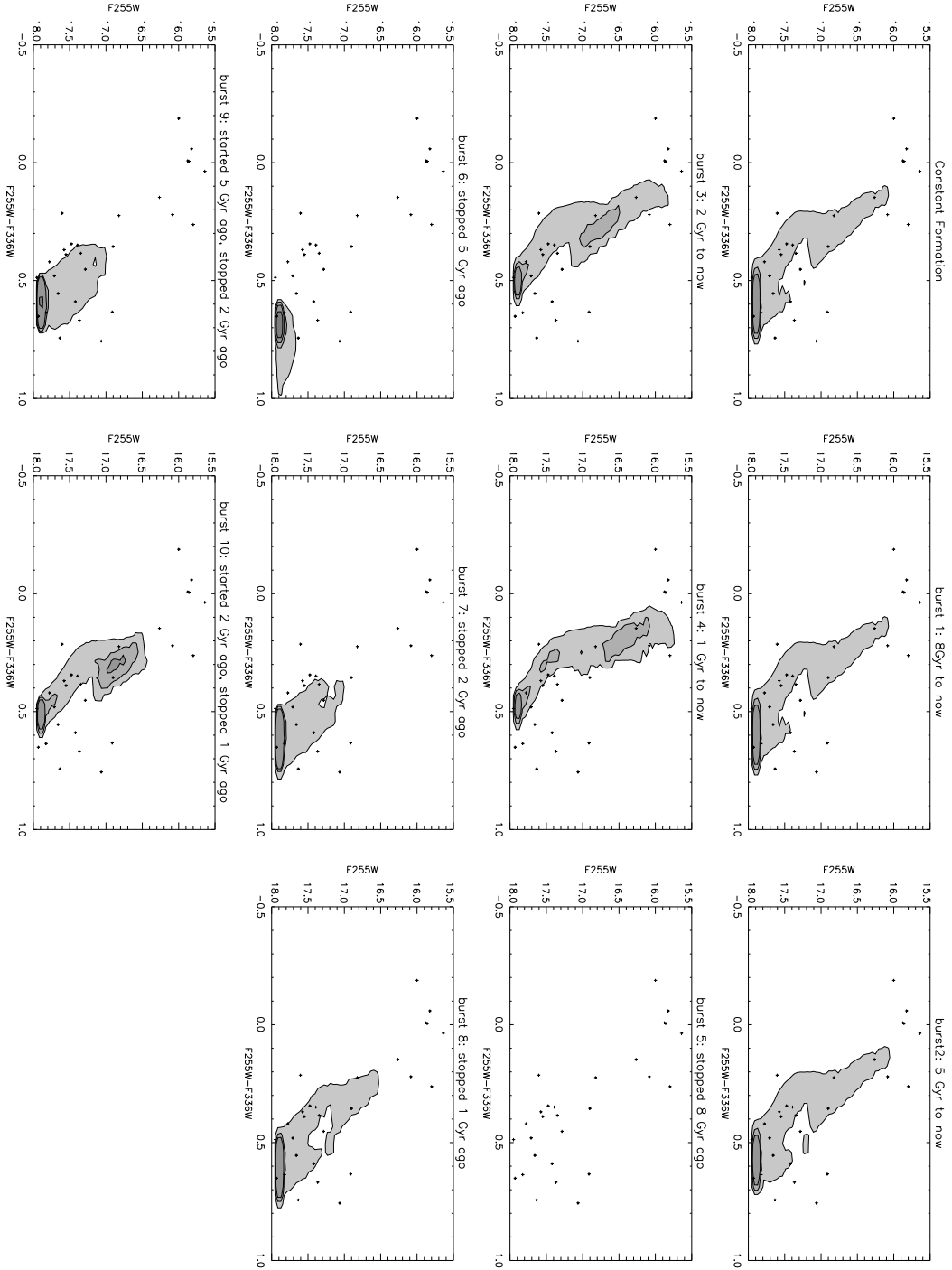} 
\caption{\label{HSTsim}Distribution of bright BSS in the
color-magnitude diagram for the HST sample, compared to theoretical collisional
models. The observations are plotted as crosses, while the grayscale contours
give the theoretical distributions, with darker colors indicating more BSS.
Different panels correspond to different eras of constant BSS formation, as
indicated at the top of each panel}
\end{figure}

\begin{figure}
\epsscale{.90}
\plotone{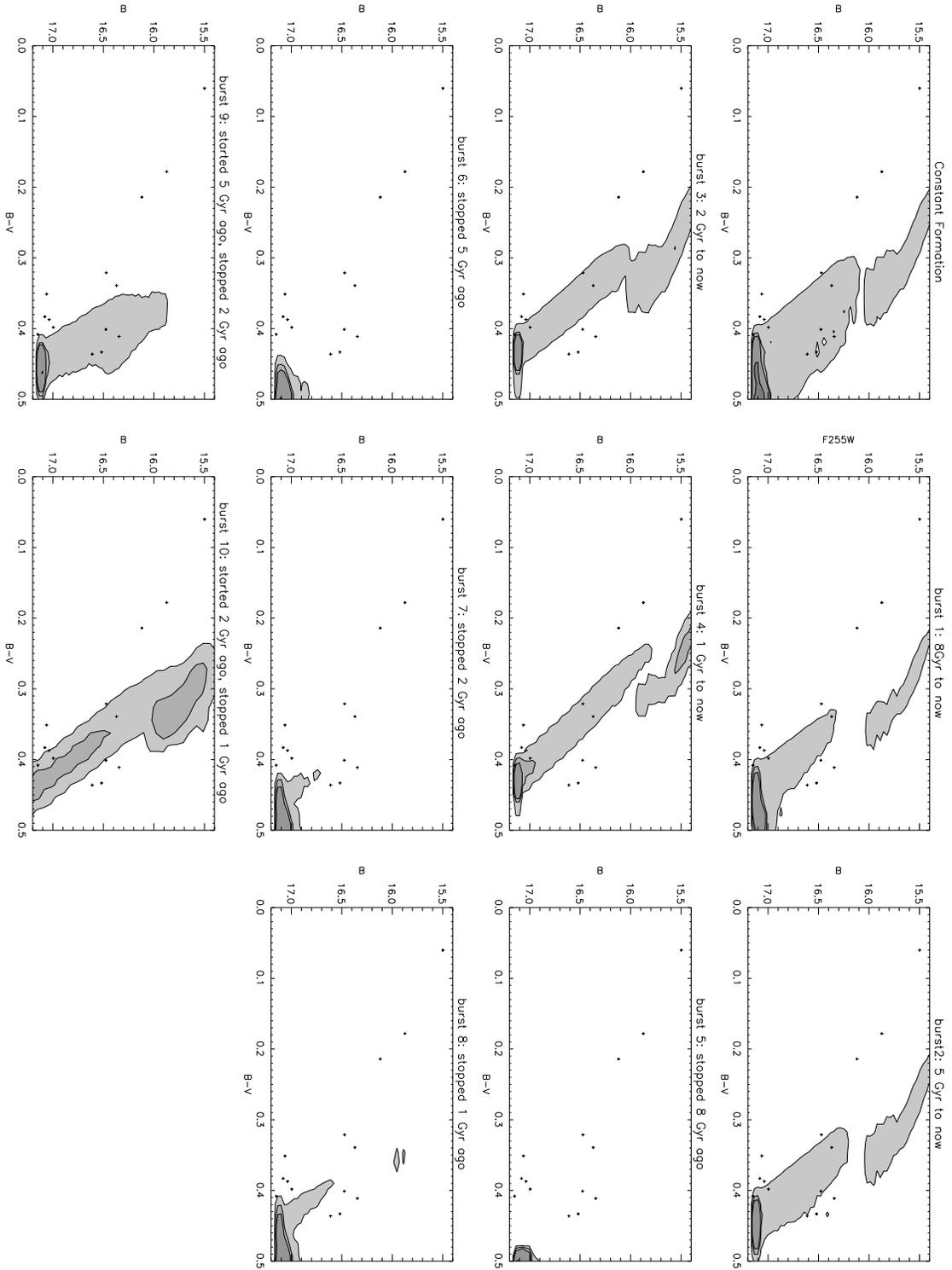}
\caption{\label{WFIsim}Same as Fig~\ref{HSTsim} for the WFI sample}
\end{figure}

We fit the models to the two sub-samples independently. The
theoretical evolutionary tracks were transformed to the ($m_{255},~
m_{255}-m_{336}$) plane or the ($B,~B-V$) plane and compared to the
data using a KS test in both luminosity and temperature. The resultant
distributions, including a variety of possible formation times, are
shown in Figures~\ref{HSTsim} and \ref{WFIsim}.

\begin{figure}
\plotone{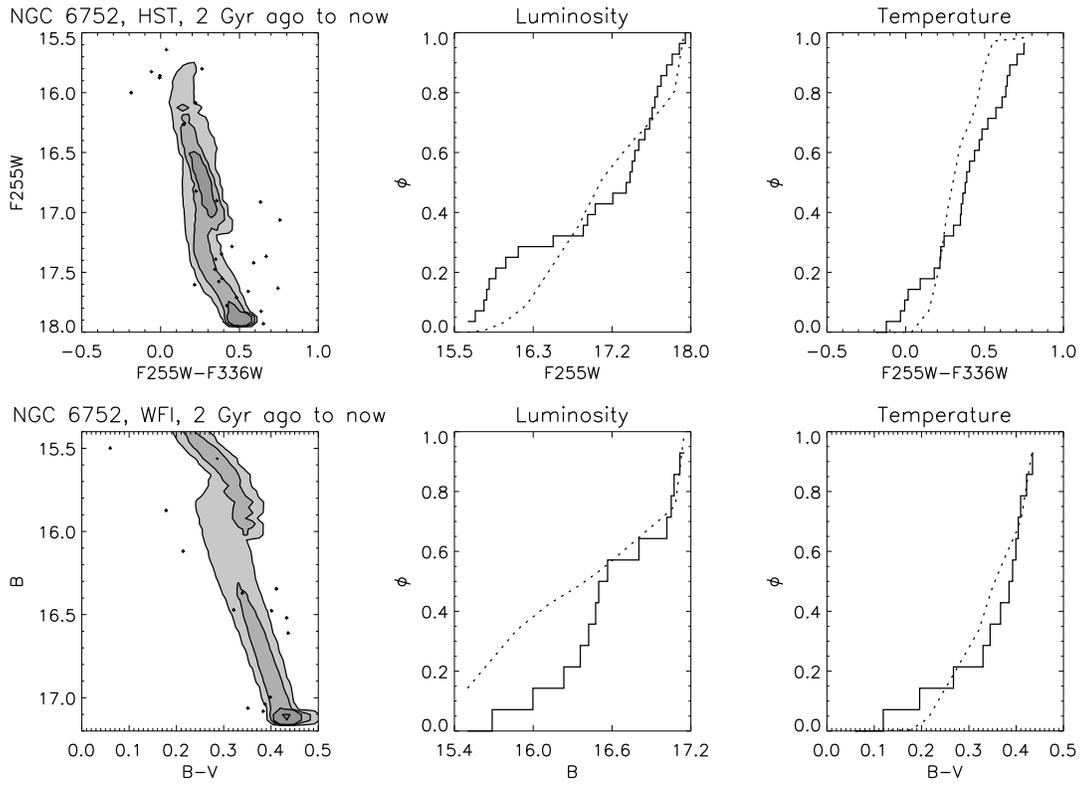}
\caption{\label{bestsim}Best fit model for both the HST and WFI samples as
determined from KS tests on the luminosity and temperature functions.}
\end{figure}

Given the sample size we must be careful not to overinterpret these
comparisons. There are several points of note: for the HST sample,
none of our models produce a significant population of the very
bright, very blue BSS which we observe. In the data there is a paucity
of stars in the range $16.2\leq m_{255} \leq 16.8$---the models tend
to populate that region as abundantly as the adjacent regions. Models
which produce BSS over a substantial part of the cluster's age predict
too many faint BSS. Indeed, as shown in Figure \ref{bestsim},
the ``best fit'' model as determined by KS tests on the luminosity
function and analogous temperature function is one in which all the
BSS were formed in the past two Gyr. The solid lines in
Figure \ref{bestsim} show the cumulative luminosity or temperature function of
the data, and the dotted lines give the theoretical predictions.

The ground-based data set is also best fit by a model in which all the
BSS were formed in the last 2 Gyr. However, it is also
plagued by similar problems in the details. While in this case we
actually over-predict the number of bright BSS, our models
are much too red to actually fit the distribution well.

In both the inner and outer regions of the cluster, the biggest
problem with our collisional models is their inability to predict
enough bright BSS compared to the fainter BSS. In addition
the luminous BSS are too blue. In one respect NGC~6752 is very much
unlike the 7 previous clusters to which we have applied this
technique.  It is best fit by a model where BSS formation {\em
started} 2 Gyr ago. All previously studied clusters, with the
exception of the low density cluster NGC 288, were best fit by models
in which the BSS {\em stopped} forming 1 to 2 Gyr ago. M3
and M92 have BSS even more luminous than those in NGC~6752. 47 Tuc
also appears to have BSS more luminous than those in NGC~6752 but this
is less certain because of the different photometric systems. A few
BSS in M80 are as luminous as those in NGC~6752, but they make up a
much smaller fraction of the BSS population. In M3, M92, \& M80, the
BSS are not as hot as in NGC~6752 and in that respect do not present a
challenge to the models. We can achieve much better formal fits to the
BSS magnitude distributions in M3, M80, and M92 because much of the
weight of those fits comes from the less luminous BSS. The best
fitting models do not include the bright BSS present in the
observations.

We must question our assumptions about the formation mechanism for the
most luminous BSS, in general. In the case of NGC~6752 we
must also address the issue of why these stars are so hot. It is
possible that these BSS are not the product of two stars
colliding, but could actually be the product of {\em three} stars
coming together. This has been suggested for the brightest blue
straggler in NGC 6397, based on STIS spectra which suggest a mass
greater than twice the TO mass for that particular star
\citep{shara02}.  Alternatively, the BSS could be
chemically very different from our assumptions. Based on SPH
simulations of collisions between MS stars and the
subsequent evolution of non-rotating collision products
\citep{sills97, sills02}, we produce BSS evolutionary
tracks which have no mixing of hydrogen to the core or helium to the
surface. These evolutionary tracks have truncated MS
lifetimes, particularly for the more massive BSS, and
spend much of their time in the redder part of the CMD. However, if
there is some mixing mechanism present, such as stellar rotation
\citep{sills01}, then the stars can be significantly bluer, brighter,
and have longer MS lifetimes. Finally, it is possible that
the BSS were not formed through collisions at all, but
rather through the merger of the two components of a primordial binary
system. There are currently very few detailed models of a merger
between two MS stars in a binary system that could be used
as starting models for stellar evolution calculations, so we have very
little information about the evolutionary tracks of such a blue
straggler. There are arguments \citep[e.g][]{bp95} which suggest that
a binary merger should involve more mixing of helium to the surface,
producing a bluer BSS.

\section{Summary and Discussion}
\label{discussion}

We have surveyed the entire radial extension of NGC~6752 for BSS using
HST UV photometry in the center and ground based ESO-WFI photometry in the
outer parts. Our sample should be relatively complete and devoid of blends
which mimic BSS. The relative frequency of BSS compared to either HB or RGB
stars has a radial distribution which is bimodal, peaking in the center and
rising up in the outer regions of the cluster. Only two other clusters, M3 \&
47~Tuc, have surveys of similar quality, and they both have bimodal
distributions. Clearly bimodal distributions are likely to be a fairly common
feature of cluster BSS populations. More BSS surveys, covering the full spatial
extent of the host cluster, are necessary to determine just how common
bimodality is.

When compared to other clusters the specific frequency of BSS in the central
regions of NGC~6752 is found to be quite low. The current binary fraction in the
core of NGC 6752 was measured to be large (between 16 and 38\%) by
\citet{rubenstein97}, and then dropping to less than 15\% outside one core
radius. As noted in the introduction, there are many other indications of a
substantial binary population: CVs, MSPs, X-ray sources. Moreover NGC~6752 has
a high central density. A large binary population and high stellar density
should lead to efficient production of BSS, yet we observe the reverse.

We have argued that a cluster's dynamic state might be important.  For
example, M80 has very large BSS population and maybe at the onset of
core contraction \citep{ferraro99b}. However our own models of M80
suggest that the BSS population may not be linked to core
contraction. The models suggest that BSS formation
ended a few Gyr ago---too long ago to have the BSS formation
tightly linked to core collapse, especially if the core collapse is
happening now. Elsewhere, we have argued \citep{ferraro03b} that the
core of NGC~6752 is undergoing a post-collapse bounce, i.e., is in a
more advanced dynamical state than M80. Maybe the grand epoch for BSS
in NGC~6752 was long in the past?  Our modeling indicates
otherwise---the large number of luminous BSS compared to fainter ones
suggests recent BSS formation.

The relatively large number of luminous and hot BSS in the cluster
core casts some doubt on our model formation mechanism. A
formation mechanism which occurs favorably in an environment with
many binary stars could work better in the core of this
cluster. Both the triple collision model and the binary merger model
should be considered in more detail for the HST observations of BSS in
NGC 6752. The ground-based data, on the other hand, may simply be
better fit by more realistic models of BSS (perhaps
involving some rotation).

\acknowledgements

We  warmly thank Paolo Montegriffo for assistance during the astrometry
procedure.  The financial  support of the  Agenzia Spaziale Italiana (ASI)   
and the MIUR (Ministero dell' Istruzione, dell' Universit\`a e della Ricerca)
is kindly acknowledged. This research has made use of  the GSCII catalog which 
has been produced by the Space Telescope Science Institute and the Osservatorio
Astronomico di Torino and of  the ESO/ST-ECF Science Archive facility which is
a joint collaboration of  the European Southern Observatory and the Space
Telescope - European Coordinating Facility. RTR is partially supported by STScI
grant GO-8709 and NASA LTSA grant NAG 5-6403.

\clearpage

\clearpage

\clearpage

\clearpage

\clearpage

\clearpage

\clearpage

\clearpage

\clearpage

\clearpage

\begin{deluxetable}{lcccccc}
\tablewidth{15.5cm}  
\tablecaption{\label{tab1}The BSS population in NGC~6752-HST sample}
\startdata \\
\hline
\hline
   &  &    &   &   &   &   \\ 
 Name  & Identification &  m$_{255}$ & m$_{336}$ & m$_{555}$ & Ra & Dec \\ 
   &  &    &   &   &   &   \\ 
\hline        
BSS-HST 1         & 100547 & 15.64 & 15.61 & 15.28 & 287.7177960 & -59.9854470 \\ 
BSS-HST 2         & 302241 & 15.80 & 15.54 & 15.42 & 287.7669809 & -59.9854152 \\ 
BSS-HST 3$^\ast$  & 100544 & 15.83 & 15.88 &  0.00 & 287.7180070 & -59.9857744 \\ 
BSS-HST 4         & 100879 & 15.86 & 15.87 & 15.57 & 287.7168449 & -59.9846164 \\ 
BSS-HST 5         & 200439 & 15.88 & 15.88 & 15.77 & 287.7133298 & -59.9727028 \\ 
BSS-HST 6 	  & 101104 & 16.00 & 16.19 & 15.76 & 287.7226674 & -59.9790308 \\ 
BSS-HST 7 	  & 201881 & 16.09 & 15.87 & 15.59 & 287.7411119 & -59.9633487 \\ 
BSS-HST 8 	  & 102301 & 16.27 & 16.12 & 15.81 & 287.7142228 & -59.9783230 \\ 
BSS-HST 9         & 100145 & 16.82 & 16.60 & 16.23 & 287.7196979 & -59.9859308 \\  
BSS-HST 10 	  & 400382 & 16.90 & 16.55 & 16.20 & 287.7293083 & -59.9826694 \\ 
BSS-HST 11 	  & 200769 & 16.91 & 16.28 & 15.96 & 287.7117728 & -59.9696571 \\ 
BSS-HST 12 	  & 101633 & 17.07 & 16.31 & 15.99 & 287.7148306 & -59.9829903 \\ 
BSS-HST 13 	  & 400196 & 17.29 & 16.83 & 16.56 & 287.7546718 & -59.9891209 \\ 
BSS-HST 14 	  & 200615 & 17.35 & 16.96 & 16.74 & 287.7141448 & -59.9718412 \\ 
BSS-HST 15 	  & 400824 & 17.37 & 16.70 & 16.42 & 287.7248176 & -59.9839133 \\ 
BSS-HST 16 	  & 102747 & 17.39 & 17.04 & 16.85 & 287.7173063 & -59.9840910 \\ 
BSS-HST 17 	  & 400057 & 17.42 & 16.83 & 16.60 & 287.7302078 & -59.9810638 \\ 
BSS-HST 18 	  & 100697 & 17.48 & 17.13 & 16.86 & 287.7170084 & -59.9854774 \\ 
BSS-HST 19	  & 101003 & 17.63 & 16.89 & 16.53 & 287.7211611 & -59.9805172 \\ 
\hline
BSS-HST 20 	  & 200325 & 17.55 & 17.16 & 17.02 & 287.7168190 & -59.9744878 \\ 
BSS-HST 21        & 100375 & 17.58 & 17.21 & 17.13 & 287.7172990 & -59.9858952 \\ 
BSS-HST 22$^\ast$ & 402781 & 17.61 & 17.39 &  0.00 & 287.7326068 & -59.9968094 \\ 
BSS-HST 23 	  & 402020 & 17.66 & 17.11 & 16.96 & 287.7227464 & -59.9894086 \\ 
BSS-HST 24 	  & 400970 & 17.71 & 17.23 & 16.90 & 287.7269696 & -59.9853010 \\ 
BSS-HST 25 	  & 200345 & 17.78 & 17.36 & 17.33 & 287.6964971 & -59.9686652 \\ 
BSS-HST 26 	  & 101463 & 17.83 & 17.19 & 16.90 & 287.7132449 & -59.9851350 \\ 
BSS-HST 27 	  & 100650 & 17.93 & 17.28 & 16.90 & 287.7174733 & -59.9848700 \\ 
BSS-HST 28 	  & 100538 & 17.95 & 17.47 & 17.22 & 287.7179025 & -59.9857687 \\ 
\hline
\enddata
\end{deluxetable}


\begin{deluxetable}{lccccc}
\tablewidth{14.5cm}  
\tablecaption{\label{tab2}The BSS population in NGC~6752-WFI sample}
\startdata \\
\hline
\hline
   &  &    &   &   &   \\ 
 Name  & Identification &  B & V & Ra & Dec \\ 
   &  &    &   &   &    \\ 
\hline        

BSS-WFI 1	 & 200185 & 15.50 & 15.44 & 287.79768541 & -60.01398330 \\ 
BSS-WFI 2	 & 300157 & 15.88 & 15.70 & 287.35904310 & -59.86002383 \\ 
BSS-WFI 3	 & 200384 & 16.34 & 15.89 & 287.64642963 & -59.99679754 \\
BSS-WFI 4	 & 600037 & 16.12 & 15.91 & 287.48772573 & -60.23306210 \\
BSS-WFI 5	 & 204257 & 16.35 & 15.94 & 287.78519229 & -60.01891775 \\ 
BSS-WFI 6	 & 210167 & 16.37 & 16.03 & 287.64246963 & -59.94914751 \\ 
BSS-WFI 7	 & 100153 & 16.48 & 16.08 & 287.97965536 & -59.85447178 \\ 
BSS-WFI 8	 & 204708 & 16.52 & 16.09 & 287.82375822 & -60.01162798 \\ 
BSS-WFI 9	 & 300014 & 16.47 & 16.15 & 287.55954128 & -60.08444017 \\ 
BSS-WFI 10	 & 100055 & 16.61 & 16.18 & 287.98656952 & -60.00510958 \\ 
BSS-WFI 11	 & 200063 & 17.00 & 16.60 & 287.72523309 & -60.05433501 \\ 
BSS-WFI 12	 & 207299 & 17.04 & 16.65 & 287.79184573 & -59.98062382 \\ 
BSS-WFI 13	 & 212606 & 17.08 & 16.70 & 287.59023051 & -59.89508558 \\ 
BSS-WFI 14	 & 100306 & 17.06 & 16.71 & 287.92303829 & -60.08714733 \\ 
BSS-WFI 15	 & 210826 & 17.15 & 16.74 & 287.69627737 & -59.93746201 \\ \hline
\enddata
\end{deluxetable}

\begin{deluxetable}{lccccccc} 
\tablewidth{15.5cm} 

\tablecaption{\label{comparison} Parameters of Clusters with HST Observations of BSS} 
\startdata \\
\hline\hline
\colhead{Cluster} & 
\colhead{$\log\,\rho_0$} & 
\colhead{Mass} &   
\colhead{$\sigma_0$} &
\colhead{\bbsshb} & 
\colhead{ \rhalf  } & 
\colhead{ $r_c$ } & 
\colhead{\rhalf/\rcore} \\ 
 & \colhead{\scriptsize{$[M_{\odot}\,pc^{-3}]$}}
 & \colhead{\scriptsize{$[Log(M/M_{\odot})]$}}
 & \colhead{\scriptsize{$[{\rm km\,s^{-1}}]$}}
 & ~
 & \colhead{\scriptsize(arcsec)}
 & \colhead{\scriptsize(arcsec)}
 & \colhead{} 
 \\
\colhead{(1)} & \colhead{(2)} & \colhead{(3)} & \colhead{(4)} &
\colhead{(5)} & \colhead{(6)} & \colhead{(7)} & \colhead{(8)} \\
 \hline
\small{NGC~5272(M3)}   & 3.5 & 5.8 & 5.6 &0.28 &  22\arcsec &   30\arcsec & 0.73  \\ 
\small{NGC~6205(M13)}  & 3.4 & 5.8 & 7.1 &0.07 &  46\arcsec &   40\arcsec & 1.15  \\ 
\small{NGC~6093(M80)}  & 5.4 & 6.0 &12.4 &0.44 &   7\arcsec &  6.5\arcsec & 1.07  \\ 
\small{NGC~6254(M10)}  & 3.8 & 5.4 & 5.6 &0.27 &  34\arcsec &   40\arcsec & 0.85  \\ 
\small{NGC~288}        & 2.1 & 4.9 & 2.9 &0.92 &  60\arcsec &   85\arcsec & 0.71  \\ 
\small{NGC~6341(M92)}  & 4.4 & 5.3 & 5.9 &0.33 &  15\arcsec &   14\arcsec & 1.07  \\
\small{NGC~6752}       & 5.2 & 5.2 & 4.5 &0.18 &  23\arcsec & 13.7\arcsec & 1.67  \\
                       &     &     &     &     & 5.7\arcsec &   28\arcsec & 0.82  \\
\small{NGC~104(47Tuc)} & 5.1 & 6.1 & 11.5&$\sim$ 0.2 &  16\arcsec &   21\arcsec & 0.76   \\
\enddata 
\end{deluxetable}


\begin{thebibliography}{}


\bibitem[Bailyn, 1995]{bailyn95}
Bailyn, C.~D. \ 1995, ARA\&A, 33, 133

\bibitem[Bailyn \& Pinsonneault, 1995]{bp95} 
Bailyn,  C.~D.~\& Pinsonneault, M.~H.\ 1995, \apj, 439, 705 

\bibitem[Buonanno et al., 1983]{buonanno83}
Buonanno, R.,  Buscema, G., Corsi, C.E., Ferraro, I., \& Iannicola, G. \
 1983, A\&A, 126, 278 

\bibitem[Buonanno et al., 1986]{buonanno86}
Buonanno, R., Corsi, C.~E., Iannicola, G., \& Fusi Pecci, F. \ 1986, A\&A, 159,
189

\bibitem[Buonanno \& Iannicola, 1989]{buonanno89}
Buonanno, R., \&  Iannicola, G. \ 1989, PASP, 101, 294 

\bibitem[Colpi Possenti \& Gualandris, 2002]{colpi02}
Colpi, M., Possenti, A., \& Gualandris, A. \ 2002, \apj, 570, L85

\bibitem[Colpi Mapelli \& Possenti, 2003]{colpi03}
Copli, M., Mapelli, M., \& Possenti, A \ 2003, \apj, 599, 1260

\bibitem[D'Amico et al., 2002]{damico02}
D'Amico, N., Possenti, A., Fici, L., Manchester, R.~N., Lyne, A.~G., Camilo, F.,
\& Sarkissian, J. \ 2002, \apj, 570, L89


\bibitem[Dorman Rood \& O'Connell, 1995]{dorman95}
Dorman, B., Rood, R.~T., O'Connell, R. \ 1995, \apj, 442, 105

\bibitem[Ferraro et al., 1993]{ferraro93}
Ferraro, F.~R., Fusi Pecci, F., Cacciari, C., Corsi, C., Buonanno, R., Fahlman,
G.~G., \& Richer, H.~B. \ 1993, \aj, 106, 2324

\bibitem[Ferraro Fusi Pecci \& Bellazzini, 1995]{ferraro95}
Ferraro, F.~F, Fusi Pecci, F., \& Bellazzini, M. \ 1995 A\&A, 294, 80

\bibitem[Ferraro et al., 1997]{ferraro97} 
Ferraro, F.R., Paltrinieri, B., Fusi Pecci, F., Cacciari, C., Dorman, B., Rood,
R.T., Buonanno, R., Corsi, C.E.,  Burgarella, D., Laget, M. \ 1997,  A\&A, 324,
915.

\bibitem[Ferraro et al., 1999a]{ferraro99}
Ferraro, F.~R., Paltrinieri, B., \& Cacciari, C. \ 1999a Mem SAIt, 70, 599

\bibitem[Ferraro et al., 1999b]{ferraro99b}
Ferraro, F.~R., Paltrinieri, B., Rood, R.~T., \& Dorman, B. \ 1999b, \apj, 522, 983

\bibitem[Ferraro et al., 2001]{ferraro01} 
Ferraro, F.R., D'Amico, N., Possenti, A., Mignani, R., \& Paltrinieri, B. \ 2001,
\apj, 561, 337

\bibitem[Ferraro et al., 2003a]{ferraro03a}
Ferraro, F.~R., Possenti, A., Sabbi, E., Lagani, P., Rood, R.~T., D'Amico, N., \&
Origlia, L \ 2003a, \apj, 595, 464

\bibitem[Ferraro et al., 2003b]{ferraro03b}
Ferraro, F.~R., Sills, A., Rood, R.~T., Paltrinieri, \& B., Buonanno, R. \ 2003b,
\apj, 588, 464

\bibitem[Ferraro et al., 2004]{ferraro03c}
Ferraro, F.~R., Beccari, G., Rood, R.~T., Bellazzini, M., Sills, A., Sabbi, E. \
2004, \apj, 603,127

\bibitem[Fusi Pecci et al., 1992]{fusipecci92}
Fusi Pecci, F., Ferraro, F.~R., Corsi, C.~E., Cacciari, C., \& Buonanno, R. \ 
1992, \aj, 104, 1831

\bibitem[Guenther et al., 1992]{guenther92}
Guenther, D.~B., Demarque, P., Kim, Y.~C., \& Pinsonneault, M.~H. \ 1992, \apj,
387, 372

\bibitem[Harris, 1996]{harris96}
Harris, W.E. 1996, AJ, 112, 1487

\bibitem[Hills \& Day, 1976]{hills76}
Hills, J.~G., \& Day, C.~A. \ 1976 ApJ, 17, L87

\bibitem[Holtzmann et al., 1995]{holtzmann95}
Holtzmann, J.~A., Burrows, C.~J., Casertano, S., Hester, J.~J., Trauger, J.~T.,
Watson, A.~M., \& Worthey, G. 1995, PASP, 107, 1065

\bibitem[McMillan \& Hut, 1996]{mcmillan96}
McMillan, S.~L.~W., \& Hut, P. \ 1996, \apj, 467, 348

\bibitem[Meylan \& Heggie, 1997]{meylan97}
Meylan, G., \& Heggie, D.~G. \ 1997, A\&A Rev., 8, 1

\bibitem[Pryor \& Meylan, 1993]{pryor93}
Pryor, C., \& Meylan, G. 1993, in ASP Conf. Ser. 50, Structure and Dynamics of
Globular Clusters, ed. S.~G. Djorgovski \& G. Meylan (San Francisco: ASP), 357

\bibitem[Pooley et al., 2002]{pooley02}
Pooley, B., et al. \ 2002 \apj, 569, 405

\bibitem[Rubenstein \& Bailyn(1997)]{rubenstein97} 
Rubenstein,  E.~P.~\& Bailyn, C.~D.\ 1997, \apj, 474, 701 

\bibitem[Sandage, 1953]{sandage53} 
Sandage, A.~R. \ 1953, \aj, 58, 61 

\bibitem[Sills \& Lombardi, 1997]{sl97}
Sills, A., Lombardi, J.~C.~Jr. \ 1997, \apj, 484, L51

\bibitem[Sills et al., 1997]{sills97}
Sills, A., Lombardi,  J.~C., Bailyn, C.~D., Demarque, P., Rasio, F.~A., \&
Shapiro, S.~L.\ 1997,  \apj, 487, 290 

\bibitem[Sills \& Bailyn, 1999]{sills99}
Sills, A., Bailyn, C.~D. \ 1999, \apj, 513, 428

\bibitem[Sills et al., 2000]{sills00}
Sills, A., Bailyn, C.~D., Edmonds, P.~D., Gilliland, R.~L. \ 2000, \apj, 535, 298

\bibitem[Sills et al., 2001]{sills01}
Sills, A., Faber, J.~A.,  Lombardi, J.~C., Rasio, F.~A., \& Warren, A.~R.\
2001, \apj, 548, 323

\bibitem[Sills et al., 2002]{sills02} 
Sills,  A., Adams, T., Davies, M.~B., \& Bate, M.~R.\ 2002, \mnras, 332, 49 

\bibitem[Shara, 2002]{shara02}
Shara, M.~M.\ 2002, ASP 
Conf.~Ser.~263: Stellar Collisions, Mergers and their Consequences





\end{thebibliography}
\end{document}